\documentclass[letterpaper,12pt]{article}
\usepackage{array}
\usepackage{dcolumn}
\usepackage{float}
\usepackage{natbib}
\usepackage{graphicx}
\usepackage{color}
\usepackage{amsmath}
\usepackage{amsfonts}
\usepackage{amsopn}
\usepackage{amsthm}
\usepackage{lscape}
\usepackage{multirow}

\theoremstyle{definition}

\theoremstyle{remark}
\newtheorem{rmk}{Remark}

\DeclareMathOperator{\pr}{P}
\DeclareMathOperator{\epn}{E}
\DeclareMathOperator{\var}{var}

\DeclareMathOperator{\expit}{expit}

\newcommand{\bmX}{\boldsymbol X}
\newcommand{\bmO}{\boldsymbol O}
\newcommand{\bmx}{\boldsymbol x}
\newcommand{\bmd}{\boldsymbol d}
\newcommand{\bmzero}{\boldsymbol 0}
\newcommand{\bmbeta}{\mbox{\boldmath${\beta}$}}
\newcommand{\bmgamma}{\mbox{\boldmath${\gamma}$}}
\newcommand{\bmpsi}{\mbox{\boldmath${\psi}$}}

\renewcommand{\baselinestretch}{1.8}

\begin{document}
\begin{center}
{\LARGE Outcome Regression Methods for Analyzing Hybrid Control Studies: Balancing Bias and Variability}
\end{center}

\vspace{0.5cm}
\begin{center}
Zhiwei Zhang$^{1,*}$, Jialuo Liu$^1$, and Wei Liu$^2$\\
$^1$Biostatistics Innovation Group, Gilead Sciences, Foster City, California, USA\\
$^2$School of Management, Harbin Institute of Technology, Harbin, China\\
$^*$zhiwei.zhang6@gilead.com
\end{center}

\vspace{0.5cm}
\centerline{\bf Abstract}

There is growing interest in a hybrid control design in which a randomized controlled trial is augmented with an external control arm from a previous trial or real world data. Existing methods for analyzing hybrid control studies include various downweighting and propensity score methods as well as methods that combine downweighting with propensity score stratification. In this article, we describe and discuss methods that make use of an outcome regression model (possibly in addition to a propensity score model). Specifically, we consider an augmentation method, a G-computation method, and a weighted regression method, and note that the three methods provide different bias-variance trade-offs. The methods are compared with each other and with existing methods in a simulation study. Simulation results indicate that weighted regression compares favorably with other model-based methods that seek to improve efficiency by incorporating external control data. The methods are illustrated using two examples from urology and infectious disease. 

\vspace{.5cm}
\noindent{Key words:}
augmentation; covariate adjustment; double robustness; external control; G-computation; weighted regression

\section{Introduction}\label{intro}

Randomized controlled trials (RCTs) are widely considered the gold standard for evaluating the safety and effectiveness of medical treatments. Randomization helps to balance observed and unobserved baseline covariates between treatment groups and provides a solid basis for (asymptotically) unbiased estimation of treatment effects. When randomization is infeasible, a practical alternative is to conduct a single-arm study of the experimental treatment and compare the results with an external control group, which may be available from a previous study or real world data \citep{fda23}. In recent years, a hybrid control design in which an RCT is augmented with external control data has drawn a great deal of attention. Potential benefits of the hybrid control design (compared to the RCT-only design) include higher precision and power, lower cost, and easier enrollment (owing to a more favorable allocation ratio in the RCT). On the other hand, the external control group may differ systematically in important patient characteristics from the RCT population. Such differences may result in estimation bias and type I error inflation if they are not accounted for appropriately.

One way to mitigate the impact of possible differences between the internal and external control groups is downweighting external control subjects before combining them with internal control subjects. Downweighting is typically done in the Bayesian framework using a power prior \citep{ic00,n10} but can also be conducted in a frequentist manner \citep[e.g.,][]{t22}. The weight parameter (i.e., the power in a power prior) may be difficult to specify prospectively but can be chosen adaptively using a Bayesian hierarchical model \citep{h11,v14}, an empirical Bayes method \citep{gh17}, or a frequentist method \citep{t22}. In essence, the adaptive downweighting approach adapts the weight parameter to the observed outcome data in the internal and external control groups in such a way that larger differences between the two groups generally lead to more severe downweighting.

Another general approach to the hybrid control problem is to cast the problem as a causal inference problem and borrow ideas from the causal inference literature \citep[e.g.,][]{r86,rr84,rr85,r00,vr03,br05,vr11}. A key assumption for this approach is that control outcomes are exchangeable between internal and external control subjects upon conditioning on a set of baseline covariates that are measured in both the RCT and the external control data. This exchangeability assumption is analogous to the assumption of no unmeasured confounders in observational studies, and allows us to borrow information from external control data for improved efficiency in treatment effect estimation. Several causal inference methods have been adapted to the hybrid control problem, and they are typically based on a model for the propensity score \citep[PS;][]{rr83}, which may be defined as the conditional probability (based on covariate values) that a control subject in the hybrid control study belongs to the RCT. Estimated PS values can be used to match each internal control subject with one or more external control subjects, stratify the combined control group so that the internal and external control subjects are more comparable within each stratum, or assign a weight proportional to $\text{PS}/(1-\text{PS})$ to each external control subject. Most recently, various combinations of propensity score and downweighting methods have also been suggested \citep[e.g.,][]{w19,c20,lu22,f23,w23}.

The causal inference literature also contains methods that make use of an outcome regression (OR) model (possibly in addition to a PS model), including the G-computation method \citep{r86} and various doubly robust methods \citep[e.g.,][]{vr03,br05,vr11}. These methods have desirable properties in terms of asymptotic bias and variance but, as far as we know, have not yet been utilized in the hybrid control setting. In this article, we describe and discuss several methods for the hybrid control problem that leverage an OR model with or without a PS model. Specifically, we consider an augmentation method that is increasingly adopted for covariate adjustment, a simple G-computation method from \citet{r86}, and a weighted regression method considered by \citet{hi01} and others in different contexts. The first two methods do not require a PS model, while the last one does. We note that the three methods provide different bias-variance trade-offs when the possibility of model mis-specification is taken into account. The augmentation method is protected against mis-specification bias and has limited capacity for efficiency improvement. The G-computation method is able to achieve higher efficiency by incorporating external control data when the OR model is correct, and is subject to bias when the OR model is mis-specified. The weighted regression method is doubly robust in the sense of remaining consistent if one or both of the OR and PS models are correct, and is similar to the G-computation method in its capacity for efficiency improvement. The methods are compared with each other and with some existing methods (including a PS weighting method) in a simulation study. Simulation results indicate that weighted regression compares favorably with other model-based methods that seek to improve efficiency by incorporating external control data.

The rest of the article is organized as follows. In the next section, we formulate the estimation problem and describe the three OR-based methods. In Section \ref{sim}, we report a simulation study that compares the three methods and some existing methods. Section \ref{ex} presents two illustrative examples, and Section \ref{disc} gives concluding remarks. Technical proofs are provided in Supplementary Materials.

\section{Methodology}\label{meth}

\subsection{Preliminaries}\label{pre}

For a generic subject in a hybrid control study, let $Z$ be an indicator for the source cohort (1 RCT; 0 external), $\bmX$ a vector of baseline covariates, $A$ a treatment indicator (1 experimental; 0 control), and $Y$ a clinical outcome of interest. The combined study data will be conceptualized as independent copies of $\bmO=(Z,\bmX,A,Y)$ and denoted by $\bmO_i=(Z_i,\bmX_i,A_i,Y_i)$, $i=1,\dots,n$.

For each $a\in\{0,1\}$, let $Y(a)$ denote the potential outcome for treatment $a$ arising from an individual patient, and let $\mu_a=\epn\{Y(a)|Z=1\}$, which represents the mean outcome for treatment $a$ in the RCT population. Common measures of the experimental treatment effect (relative to the control treatment) include the mean difference $\mu_1-\mu_0$, the log mean ratio $\log(\mu_1/\mu_0)$ for outcomes with positive means, and the log odds ratio $\log[\mu_1(1-\mu_0)/\{\mu_0(1-\mu_1)\}]$ for binary outcomes. Each of these can be written as $\delta=g(\mu_1)-g(\mu_0)$, where $g$ is, respectively, the identity function, the log function, or the logit function. In general, $g$ can be any smooth and strictly increasing function specified by the investigator.

To connect the causal estimand $\delta$ with the observed data, we note that the hybrid control design implies that $A$ is independent of all baseline variables within the RCT; in particular,
$$
\pr\{A=1|Z=1,\bmX,Y(0),Y(1)\}=\pr(A=1|Z=1)=:\pi\in(0,1).
$$
This allows $\mu_a$, $a\in\{0,1\}$, to be identified as $\mu_a=\epn(Y|Z=1,A=a)$ and estimated using the RCT data only. The external control data provide an opportunity to improve efficiency in estimating $\mu_0$ and $\delta$.

In the hybrid control design we consider, only the control arm of the RCT is augmented with external data; that is, $\pr(A=0|Z=0)=1$. The methods and observations in this article extend easily to hybrid study designs where both arms of an RCT are augmented with external data \citep[e.g.,][]{li22}, though we do not explicitly consider such designs in this article.

A key assumption that allows us to borrow information from external control data is the mean exchangeability assumption:
\begin{equation}\label{exch}
\epn(Y|\bmX,A=0,Z=1)=\epn(Y|\bmX,A=0,Z=0)=:m_0(\bmX).
\end{equation}
The function $m_0$ will be referred to as the OR function for estimating $\mu_0$. Assumption \eqref{exch} allows the mean control outcome to differ between the RCT and the external study, provided the difference can be explained by the covariates in $\bmX$. Indeed, under assumption \eqref{exch}, $\mu_0=\epn(Y|A=0,Z=1)=\epn\{m_0(\bmX)|Z=1\}$ may differ from $\epn(Y|A=0,Z=0)=\epn\{m_0(\bmX)|Z=0\}$ if $\bmX$ is differentially distributed between the RCT ($Z=1$) and the external study ($Z=0$). Such a difference would introduce a bias in the downweighting approach (unless the weight is fixed at 0) but can be handled easily by methods that adjust for covariates, including the propensity score methods reviewed earlier and the OR-based methods to be described shortly.

As a technical remark, the propensity score methods typically require a stronger exchangeability assumption than \eqref{exch}. In the context of causal inference with a non-randomized treatment, those methods generally rely on the strongly ignorable treatment assignment assumption of \citet{rr83}. In the present context, strong ignorability requires that $Y$ and $Z$ be conditionally independent given $\bmX$ and $A=0$, which may be written as
$$
\pr(Y\le y|\bmX,A=0,Z=1)=\pr(Y\le y|\bmX,A=0,Z=0),\qquad y\in\mathbb R.
$$
This is a distributional exchangeability assumption, which is clearly stronger than the mean exchangeability assumption \eqref{exch}.

The statistical question is how to estimate $(\mu_0,\mu_1,\delta)$ using the observed data under the assumptions stated above. Because the external control data are not informative of $\mu_1$, $\mu_1$ can be estimated using an existing estimator based on the RCT data alone. Therefore, we will focus on estimation of $\mu_0$ at first before considering estimation of $(\mu_1,\delta)$ in Section \ref{est.delta}. It is straightforward to estimate $\mu_0$ with the average outcome in the internal control arm:
$$
\overline Y_0=\frac{\sum_{i=1}^nZ_i(1-A_i)Y_i}{\sum_{i=1}^nZ_i(1-A_i)}.
$$
It is easy to see that $\overline Y_0$ is unbiased and consistent for $\mu_0$ and asymptotically normal. However, $\overline Y_0$ does not make use of the external control data or the internal covariate data, and thus may be inefficient. In the next three subsections, we consider ways to improve efficiency over $\overline Y_0$.

\subsection{Augmentation}\label{aug}

Originally developed for missing data and covariate adjustment \citep[e.g.,][]{t06,t08}, augmentation is an effective way to incorporate covariate information into the estimation of $\mu_0$. In the present context, an augmented estimator of $\mu_0$ may be obtained as
\begin{equation}\label{est.aug}
\widehat\mu_0^{\text{aug}}=\overline Y_0-\frac{\sum_{i=1}^nZ_i(1-A_i)\widehat m_0(\bmX_i)}{\sum_{i=1}^nZ_i(1-A_i)}+\frac{\sum_{i=1}^nZ_i\widehat m_0(\bmX_i)}{\sum_{i=1}^nZ_i},
\end{equation}
where $\widehat m_0(\cdot)$ is a generic estimator of $m_0(\cdot)$.

Under Assumption \eqref{exch}, the OR function $m_0$ can be estimated from the combined outcome and covariate data from all control subjects (internal and external). To fix ideas, suppose a generalized linear model is specified as a working model for $m_0$:
\begin{equation}\label{or.mod}
m_0(\bmx)=h((1,\bmx')\bmbeta),
\end{equation}
where $h$ is a specified inverse link function and $\bmbeta$ an unknown parameter vector. Then $\bmbeta$ may be estimated by solving the estimating equation
\begin{equation}\label{est.eq}
\sum_{i=1}^n(1-A_i)w^{1-Z_i}\left\{Y_i-h\left((1,\bmX_i')\bmbeta\right)\right\}
(1,\bmX_i')'=\bmzero,
\end{equation}
where $w\in[0,1]$ is an optional weight that can be used to control the impact of external data. For example, $w$ might be set to $n^*/\sum_{i=1}^n(1-Z_i)$, where $n^*$ is the effective sample size of the external control arm, which may be specified by a regulatory agency. Let $\widehat\bmbeta$ denote the solution (in $\bmbeta$) to the above equation; then we can set $\widehat m_0(\bmX_i)=h((1,\bmX_i')\widehat\bmbeta)$ in equation \eqref{est.aug}.

As an augmented estimator, $\widehat\mu_0^{\text{aug}}$ is expected to be robust against mis-specification of the working model. Regardless of the (in)correctness of the working model \eqref{or.mod}, we assume that $\widehat\bmbeta$ converges in probability to some $\bmbeta^*$. If the working model happens to be correct, $\bmbeta^*$ is the true value of $\bmbeta$ in the model. Let $m_0^*(\bmx)=h((1,\bmx')\bmbeta^*)$ and $\mu_0^*=\epn\{m_0^*(\bmX)|Z=1\}$. Clearly, $\widehat\mu_0^{\text{aug}}$ converges in probability to
$$
\mu_0-\epn\{m_0^*(\bmX)|Z=1,A=0\}+\epn\{m_0^*(\bmX)|Z=1\}=\mu_0-\mu_0^*+\mu_0^*=\mu_0,
$$
where the two conditional expectations are equal because of randomization. Under standard regularity conditions, we show in Appendix A that $\sqrt n(\widehat\mu_0^{\text{aug}}-\mu_0)$ converges to a normal distribution with mean 0 and variance
$$
\sigma_{\text{aug}}^2=\var\left(\frac{Z[(1-A)(Y-\mu_0)+(A-\pi)\{m_0^*(\bmX)-\mu_0^*\}]}{\tau(1-\pi)}\right),
$$
where $\tau=\pr(Z=1)$. Note that $\sigma_{\text{aug}}^2$ depends on $\widehat m_0$ only through its limit $m_0^*$, regardless of the variability of $\widehat m_0$ or $\widehat\bmbeta$. In Appendix A, we also show that $\sigma_{\text{aug}}^2$ is minimized when $m_0^*=m_0$, which is expected to hold if the working model is correct.

Although $\widehat\mu_0^{\text{aug}}$ is model-robust in the sense of remaining consistent and asymptotically normal when the working OR model is mis-specified, it has limited ability to improve efficiency by incorporating external control data. Recall that the minimum of $\sigma_{\text{aug}}^2$ is attained when the working OR model is correct. However, if the working OR model is correct, the same minimum can be attained without incorporating external control data. To see this, we can simply set $w=0$ in equation \eqref{est.eq} and note that $\widehat\bmbeta$ based on the internal control data alone would remain consistent for the true value of $\bmbeta$, which implies $m_0^*=m_0$ (and thus minimizes $\sigma_{\text{aug}}^2$). If the working OR model is incorrect, $\bmbeta^*$ may change when $w$ moves from 0 to a positive value, but it seems unclear that a positive $w$ will necessarily lead to a smaller value of $\sigma_{\text{aug}}^2$. Thus, although the augmentation approach allows us to incorporate external control data without introducing bias, it does not necessarily produce an efficiency advantage as a result of incorporating external control data.

\subsection{G-Computation}\label{gc}

The identity $\mu_0=\epn\{m_0(\bmX)|Z=1\}$ motivates a simple estimator of $\mu_0$:
$$
\widehat\mu_0^{\text{gc}}=\frac{\sum_{i=1}^nZ_i\widehat m_0(\bmX_i)}{\sum_{i=1}^nZ_i},
$$
where $\widehat m_0$ is a fitted OR model described in Section \ref{aug}. This can be regarded as an instance of G-computation \citep{r86}, a general approach to causal inference. Note that $\widehat\mu_0^{\text{gc}}$ corresponds exactly to the last term in $\widehat\mu_0^{\text{aug}}$.

\begin{rmk}\label{equiv}
If we exclude the external control data by setting $w=0$ in \eqref{est.eq}, the first component of the estimating equation \eqref{est.eq} then becomes
$$
\sum_{i=1}^nZ_i(1-A_i)\left\{Y_i-h\left((1,\bmX_i')\bmbeta\right)\right\}=0,
$$
which implies
$$
\overline Y_0=\frac{\sum_{i=1}^nZ_i(1-A_i)Y_i}{\sum_{i=1}^nZ_i(1-A_i)}
=\frac{\sum_{i=1}^nZ_i(1-A_i)\widehat m_0(\bmX_i)}{\sum_{i=1}^nZ_i(1-A_i)}
$$
and hence $\widehat\mu_0^{\text{gc}}=\widehat\mu_0^{\text{aug}}$. Thus, in the absence of external control data, the two estimators are typically identical. We will see that the two estimators behave quite differently after incorporating external control data.
\end{rmk}

The G-computation estimator $\widehat\mu_0^{\text{gc}}$ is expected to converge in probability to $\epn\{m_0^*(\bmX)|Z=1\}=\mu_0^*$, which equals $\mu_0$ if $m_0^*=m_0$. Thus, unlike $\widehat\mu_0^{\text{aug}}$, $\widehat\mu_0^{\text{gc}}$ is subject to bias when the working OR model is mis-specified. Under regularity conditions, $\sqrt n(\widehat\mu_0^{\text{gc}}-\mu_0^*)$ converges to a normal distribution with mean 0 and variance
$$
\sigma_{\text{gc}}^2=\var\left[\frac{Z\{m_0^*(\bmX)-\mu_0^*\}}{\tau}+\bmd(\bmbeta^*)'\bmpsi_{\widehat\bmbeta}(\bmO)\right],
$$
where $\bmd(\bmbeta^*)=\epn\{\dot h((1,\bmX)'\bmbeta^*)(1,\bmX')'|Z=1\}$, $\dot h$ is the derivative function of $h$, and $\bmpsi_{\widehat\bmbeta}(\bmO)$ is the influence function of $\widehat\bmbeta$ so that $\sqrt n(\widehat\bmbeta-\bmbeta^*)=n^{-1/2}\sum_{i=1}^n\bmpsi_{\widehat\bmbeta}(\bmO_i)+o_p(1)$. A derivation for this result is given in Appendix A.

\begin{rmk}\label{var.decomp}
Assuming that the OR model is correct (so that $m_0^*=m_0$ and $\mu_0^*=\mu_0$), we demonstrate in Appendix A that $Z\{m_0(\bmX)-\mu_0\}$ and $\bmpsi_{\widehat\bmbeta}(\bmO)$ are uncorrelated, so that
\begin{equation*}\begin{aligned}
\sigma_{\text{gc}}^2&=\var\left[\frac{Z\{m_0(\bmX)-\mu_0\}}{\tau}\right]+\var\left\{\bmd(\bmbeta^*)'\bmpsi_{\widehat\bmbeta}(\bmO)\right\}\\
&=\tau^{-1}\var\{m_0(\bmX)-\mu_0|Z=1\}+\bmd(\bmbeta^*)'\Sigma_{\widehat\bmbeta}\bmd(\bmbeta^*),
\end{aligned}\end{equation*}
where $\Sigma_{\widehat\bmbeta}=\var\{\bmpsi_{\widehat\bmbeta}(\bmO)\}$ is the asymptotic variance of $\widehat\bmbeta$. This decomposition of $\sigma_{\text{gc}}^2$ can be interpreted as follows. The first term, $\tau^{-1}\var\{m_0(\bmX)-\mu_0|Z=1\}$, is the asymptotic variance of the \lq\lq estimator" $\sum_{i=1}^nZ_im_0(\bmX_i)/\sum_{i=1}^nZ_i$ with $\widehat\bmbeta$ replaced by $\bmbeta^*$. The second term, $\bmd(\bmbeta^*)'\Sigma_{\widehat\bmbeta}\bmd(\bmbeta^*)$, represents the loss of precision due to estimating $\bmbeta^*$ with $\widehat\bmbeta$. A decrease in $\Sigma_{\widehat\bmbeta}$ in the sense of negative definiteness generally leads to a decrease in $\bmd(\bmbeta^*)'\Sigma_{\widehat\bmbeta}\bmd(\bmbeta^*)$ and hence $\sigma_{\text{gc}}^2$. Because $\Sigma_{\widehat\bmbeta}$ is expected to decrease after incorporating external control data, the G-computation approach does reward the use of external control data with reduced variability, when the working OR model is correct.
\end{rmk}

For the G-computation approach, the ideal situation would be that the OR model is correctly specified and $\widehat\bmbeta$ has little variability so that $\sigma_{\text{gc}}^2$ approaches its minimum, $\underline{\sigma}_{\text{gc}}^2:=\tau^{-1}\var\{m_0(\bmX)-\mu_0|Z=1\}$. In this ideal situation, the augmented estimator also reaches its minimum variance:
\begin{equation*}\begin{aligned}
\underline{\sigma}_{\text{aug}}^2:&=\var\left(\frac{Z[(1-A)(Y-\mu_0)+(A-\pi)\{m_0(\bmX)-\mu_0\}]}{\tau(1-\pi)}\right)\\
&=\var\left(\frac{Z[(1-A)\{Y-m_0(\bmX)\}+(1-\pi)\{m_0(\bmX)-\mu_0\}]}{\tau(1-\pi)}\right)\\
&=\var\left[\frac{Z(1-A)\{Y-m_0(\bmX)\}}{\tau(1-\pi)}\right]+\var\left[\frac{Z\{m_0(\bmX)-\mu_0\}}{\tau}\right]\\
&=\var\left[\frac{Z(1-A)\{Y-m_0(\bmX)\}}{\tau(1-\pi)}\right]+\underline{\sigma}_{\text{gc}}^2
\ge\underline{\sigma}_{\text{gc}}^2,
\end{aligned}\end{equation*}
where the decomposition is due to uncorrelatedness. Thus, under a correctly specified OR model, the G-computation approach provides an opportunity to exceed the highest level of efficiency attainable by the augmentation approach. This potential efficiency advantage comes at the expense of a potential bias due to model mis-specification.

\subsection{Weighted Regression}\label{wr}

Weighted regression refers to fitting model \eqref{or.mod} using a weighted estimating equation incorporating PS-based weights for external control subjects. The PS is defined as
$$
p(\bmX)=\pr(Z=1|\bmX)
$$
and is usually estimated using a logistic regression model such as
$$
p(\bmx)=\expit((1,\bmx')\bmgamma),
$$
where $\expit(u)=1/\{1+\exp(-u)\}$ and $\bmgamma$ is an unknown parameter vector. Let $\widehat\bmgamma$ denote the maximum likelihood estimate of $\gamma$, which solves the estimating equation
$$
\sum_{i=1}^n\left\{Z_i-\expit\left((1,\bmX_i')\bmgamma\right)\right\}
(1,\bmX_i')'=\bmzero;
$$
then $p(\bmx)$ is estimated by $\expit((1,\bmx')\widehat\bmgamma)$ and the propensity odds $p(\bmx)/\{1-p(\bmx)\}$ by $\exp((1,\bmx')\widehat\bmgamma)$. Next, we estimate $\bmbeta$ in model \eqref{or.mod} by solving the weighted estimating equation
\begin{equation*}
\sum_{i=1}^n(1-A_i)\{w^{\dagger}\exp((1,\bmX_i')\widehat\bmgamma)\}^{1-Z_i}\left\{Y_i-h\left((1,\bmX_i')\bmbeta\right)\right\}
(1,\bmX_i')'=\bmzero,
\end{equation*}
where
$$
w^{\dagger}=\frac{w\sum_{i=1}^n(1-Z_i)}{\sum_{i=1}^n(1-Z_i)\exp((1,\bmX_i')\widehat\bmgamma)}
$$
is an adjusted weight to ensure that the external control arm has the same effective sample size as in equation \eqref{est.eq}. Let $\widetilde\bmbeta$ denote the solution to the above estimating equation, and let $\widetilde m_0(\bmx)=h((1,\bmx')\widetilde\bmbeta)$. The weighted regression estimator of $\mu_0$ is given by 
$$
\widehat\mu_0^{\text{wr}}=\frac{\sum_{i=1}^nZ_i\widetilde m_0(\bmX_i)}{\sum_{i=1}^nZ_i}.
$$

Appendix A provides an asymptotic theory for $\widehat\mu_0^{\text{wr}}$ which allows the working OR and PS models to be mis-specified. We assume that $(\widetilde\bmbeta,\widetilde\bmgamma)$ converge in probability to some $(\bmbeta^{\diamond},\bmgamma^{\diamond})$. If the OR (rsp.~PS) model is correct, $\bmbeta^{\diamond}$ (rsp.~$\bmgamma^{\diamond}$) is the true parameter value in the model. In general, $\widehat\mu_0^{\text{wr}}$ converges in probability to $\mu_0^{\diamond}=\epn\{m_0^{\diamond}(\bmX)|Z=1\}$, where $m_0^{\diamond}(\bmx)=h((1,\bmx')\bmbeta^{\diamond})$. Clearly, if the OR model is correct, $m_0^{\diamond}=m_0$ and $\mu_0^{\diamond}=\mu_0$. Assuming only that the PS model is correct, it can be shown that $\mu_0^{\diamond}=\mu_0$ even though $m_0^{\diamond}$ may differ from $m_0$. Thus, $\widehat\mu_0^{\text{wr}}$ is doubly robust in the sense of being consistent for $\mu_0$ under correct specification of one or both of the OR and PS models. Regardless of model (in)correctness, $\sqrt n(\widehat\mu_0^{\text{wr}}-\mu_0^{\diamond})$ converges to a normal distribution with mean 0 and variance
$$
\sigma_{\text{wr}}^2=\var\left[\frac{Z\{m_0^{\diamond}(\bmX)-\mu_0^{\diamond}\}}{\tau}+\bmd(\bmbeta^{\diamond})'\bmpsi_{\widetilde\bmbeta}(\bmO)\right],
$$
where $\bmd(\cdot)$ is defined in Section \ref{gc} and $\bmpsi_{\widetilde\bmbeta}(\bmO)$ is the influence function of $\widetilde\bmbeta$. As in Remark \ref{var.decomp}, when the OR model is correct, $\sigma_{\text{wr}}^2$ can be decomposed as follows:
\begin{equation*}\begin{aligned}
\sigma_{\text{wr}}^2&=\var\left[\frac{Z\{m_0(\bmX)-\mu_0\}}{\tau}\right]+\var\left\{\bmd(\bmbeta^{\diamond})'\bmpsi_{\widetilde\bmbeta}(\bmO)\right\}\\
&=\tau^{-1}\var\{m_0(\bmX)-\mu_0|Z=1\}+\bmd(\bmbeta^{\diamond})'\Sigma_{\widetilde\bmbeta}\bmd(\bmbeta^{\diamond}),
\end{aligned}\end{equation*}
where $\Sigma_{\widetilde\bmbeta}=\var\{\bmpsi_{\widetilde\bmbeta}(\bmO)\}$ is the asymptotic variance of $\widetilde\bmbeta$. Thus, if $\Sigma_{\widetilde\bmbeta}$ decreases as a result of incorporating external control data, $\sigma_{\text{wr}}^2$ will decrease as well under a correct OR model.

The weighted regression method offers a different bias-variance trade-off than the augmentation and G-computation methods. Unlike $\widehat\mu_0^{\text{aug}}$, which is protected against mis-specification bais, $\widehat\mu_0^{\text{gc}}$ and $\widehat\mu_0^{\text{wr}}$ are susceptible to such bias, with $\widehat\mu_0^{\text{wr}}$ being less susceptible because of its double robustness. Thus, compared to the other two methods, weighted regression has an intermediate level of risk for bias. In terms of variability, $\widehat\mu_0^{\text{wr}}$ behaves similarly to $\widehat\mu_0^{\text{gc}}$, and both estimators are potentially able to produce a larger efficiency improvement (than is attainable by $\widehat\mu_0^{\text{aug}}$) by incorporating external control data.

\subsection{Estimation of $(\mu_1,\delta)$}\label{est.delta}

Without making additional strong assumptions, the availability of external control data is irrelevant to the estimation of $\mu_1$, which can be based on the RCT data alone. Specifically, $\mu_1$ can be estimated using an augmented estimator analogous to $\widehat\mu_0^{\text{aug}}$:
$$
\widehat\mu_1^{\text{aug}}=\overline Y_1-\frac{\sum_{i=1}^nZ_iA_i\widehat m_1(\bmX_i)}{\sum_{i=1}^nZ_iA_i}+\frac{\sum_{i=1}^nZ_i\widehat m_1(\bmX_i)}{\sum_{i=1}^nZ_i},
$$
where $\overline Y_1=(\sum_{i=1}^nZ_iA_i)^{-1}\sum_{i=1}^nZ_iA_iY_i$ and $\widehat m_1(\bmx)$ is a generic estimator of $m_1(\bmx)=\epn(Y|\bmX=\bmx,A=1,Z=1)$. For example, $\widehat m_1$ can be based on a similar regression model to \eqref{or.mod} together with an estimating equation similar to \eqref{est.eq} with $(1-A_i)w^{1-Z_i}$ replaced by $A_i$. Because the estimation of $m_1$ involves no external data, it can be shown as in Remark \ref{equiv} that $\widehat\mu_1^{\text{aug}}$ is identical to the G-computation estimator of $\mu_1$ based on the same regression model and the same estimating equation. Whether the working model for $m_1$ is correct or not, $\widehat\mu_1^{\text{aug}}$ is consistent for $\mu_1$ and asymptotically normal with asymptotic variance
$$
\var\left(\frac{Z[A(Y-\mu_1)-(A-\pi)\{m_1^*(\bmX)-\mu_1^*\}]}{\tau\pi}\right),
$$
where $m_1^*$ is the limit of $\widehat m_1$ and $\mu_1^*=\epn\{m_1^*(\bmX)|Z=1\}$. These results follow from the same arguments used to establish the analogous results for $\widehat\mu_0^{\text{aug}}$ (see Appendix A).

In general, $\delta=g(\mu_1)-g(\mu_0)$ can be estimated as $g(\widehat\mu_1)-g(\widehat\mu_0)$, where $(\widehat\mu_1,\widehat\mu_0)$ are generic estimators of $(\mu_1,\mu_0)$. The specific choices of $(\widehat\mu_1,\widehat\mu_0)$ considered here give rise to three estimators of $\delta$, whose asymptotic properties follow directly from the properties of $(\widehat\mu_1,\widehat\mu_0)$. Specifically, the estimator $\widehat\delta_{\text{aug-aug}}=g(\widehat\mu_1^{\text{aug}})-g(\widehat\mu_0^{\text{aug}})$ is consistent for $\delta$ and asymptotically normal with asymptotic variance
\begin{multline*}
\var\Bigg\{\dot g(\mu_1)\left(\frac{Z[A(Y-\mu_1)-(A-\pi)\{m_1^*(\bmX)-\mu_1^*\}]}{\tau\pi}\right)\\
-\dot g(\mu_0)\left(\frac{Z[(1-A)(Y-\mu_0)+(A-\pi)\{m_0^*(\bmX)-\mu_0^*\}]}{\tau(1-\pi)}\right)\Bigg\},
\end{multline*}
where $\dot g$ is the derivative function of $g$; the estimator $\widehat\delta_{\text{aug-gc}}=g(\widehat\mu_1^{\text{aug}})-g(\widehat\mu_0^{\text{gc}})$ converges in probability to $g(\mu_1)-g(\mu_0^*)$ and has asymptotic variance
\begin{multline*}
\var\Bigg\{\dot g(\mu_1)\left(\frac{Z[A(Y-\mu_1)-(A-\pi)\{m_1^*(\bmX)-\mu_1^*\}]}{\tau\pi}\right)\\
-\dot g(\mu_0^*)\left[\frac{Z\{m_0^*(\bmX)-\mu_0^*\}}{\tau}+\bmd(\bmbeta^*)'\bmpsi_{\widehat\bmbeta}(\bmO)\right]\Bigg\};
\end{multline*}
and $\widehat\delta_{\text{aug-wr}}=g(\widehat\mu_1^{\text{aug}})-g(\widehat\mu_0^{\text{wr}})$ converges in probability to $g(\mu_1)-g(\mu_0^{\diamond})$ and has asymptotic variance
\begin{multline*}
\var\Bigg\{\dot g(\mu_1)\left(\frac{Z[A(Y-\mu_1)-(A-\pi)\{m_1^*(\bmX)-\mu_1^*\}]}{\tau\pi}\right)\\
-\dot g(\mu_0^{\diamond})\left[\frac{Z\{m_0^{\diamond}(\bmX)-\mu_0^{\diamond}\}}{\tau}+\bmd(\bmbeta^{\diamond})'\bmpsi_{\widetilde\bmbeta}(\bmO)\right]\Bigg\}.
\end{multline*}
For all estimands $(\mu_1,\mu_0,\delta)$ and estimators, variance estimates can be obtained by replacing the $\var$ operator with sample variance and unknown parameter values with empirical estimates.

\section{Simulation}\label{sim}

We now report a simulation study that evaluates the three methods described in Section \ref{meth} in comparison to three alternative methods: an RCT-only method, an unadjusted (for covariates) downweighting method, and a PS weighting method. The RCT-only method estimates $\mu_0$ with $\overline Y_0$, the unadjusted downweightinging method estimates $\mu_0$ with $\sum_{i=1}^n(1-A_i)w^{1-Z_i}Y_i/\sum_{i=1}^n(1-A_i)w^{1-Z_i}$, and the PS weighting method estimates $\mu_0$ with
$$
\frac{\sum_{i=1}^n(1-A_i)\{w^{\dagger}\exp((1,\bmX_i')\widehat\bmgamma)\}^{1-Z_i}Y_i}{\sum_{i=1}^n(1-A_i)\{w^{\dagger}\exp((1,\bmX_i')\widehat\bmgamma)\}^{1-Z_i}}.
$$
In all three alternative methods, $\mu_1$ is estimated by $\overline Y_1$ and $\delta=g(\mu_1)-g(\mu_0)$ by substituting estimates of $(\mu_1,\mu_0)$. Recall that all three methods in Section \ref{meth} estimate $\mu_1$ with $\widehat\mu_1^{\text{aug}}$. Thus, for estimating $\mu_1$, there are only two distinct estimators ($\overline Y_1$ and $\widehat\mu_1^{\text{aug}}$) being compared.

In this simulation study, the external control arm consists of 100 subjects, and the RCT has 150 subjects, who are randomized in a 2:1 ratio to the experimental and control arms; thus $n=250$, $\tau=3/5$, and $\pi=2/3$. The covariate vector $\bmX$ may be based on one or two underlying covariates. We first describe the one-covariate setting, where $X\sim N(0,1)$ in the RCT and $X\sim N(-0.5, 1.5^2)$ in the external control arm. It follows from Bayes' law that
\begin{equation}\label{ps.mod.1}
p(X)=\pr(Z=1|X)=\expit(\gamma_0+\gamma_1X+\gamma_2X^2),
\end{equation}
for some $\bmgamma=(\gamma_0,\gamma_1,\gamma_2)'$, whose exact value is not important to know. The treatment assignment mechanism may be described as $\pr(A=1|Z,X)=\pi Z$. The outcome variable $Y$ may be binary or continuous, and follows a generalized linear model:
$$
\epn(Y|A,Z,X)=h\left(-0.5+0.3X+0.5X^2+A(0.5-0.1X)\right),
$$
where $h$ is the expit function for a binary outcome and the identity function for a continuous outcome. In the binary case, the above expression specifies the conditional distribution of $Y$ given $(A,Z,X)$. In the continuous case, $Y$ is generated as $\epn(Y|A,Z,X)+\varepsilon$, where $\varepsilon\sim N(0,1)$ independently of $(A,Z,X)$. In each case, $10^4$ sets of study data are simulated.

For both continuous and binary outcomes, the effect measure of interest is the mean/rate difference $\delta=\mu_1-\mu_0$. The true values of $(\mu_1,\mu_0,\delta)$ are $(0.5,0,0.5)$ in the continuous case and approximately $(0.604, 0.488, 0.116)$ in the binary case. The three parameters are estimated using the six methods described earlier with $w=1/2$ (so the effective sample size of the external control arm is equal to the expected sample size of the internal control arm) and with working PS and OR models that may or may not be correctly specified. The correct PS model is given by equation \eqref{ps.mod.1}, and the incorrect PS model results from omitting the quadratic term in \eqref{ps.mod.1}. The correct OR model (for each treatment) is a linear (for continuous $Y$) or logistic (for binary $Y$) regression model with $(1,X,X^2)$ as linear terms, and the incorrect one omits the term $X^2$. The augmented estimator $\widehat\mu_1^{\text{aug}}$ requires a working OR model for the experimental treatment, which is specified identically to the control OR model but estimated separately. Thus, in each method that requires working OR models for the two treatments, the two models are either both correct or both incorrect.

We now describe the setting where $\bmX$ is based on two covariates $(X_1,X_2)$. Most specifications of the one-covariate setting remain applicable, and we will focus on important differences between the two settings. The two covariates are independently distributed as $N(0,1)$ in the RCT and as $N(-0.5, 1.5^2)$ in the external control arm, which implies (by Bayes' law) that
\begin{equation}\label{ps.mod.2}
p(\bmX)=\pr(Z=1|\bmX)=\expit(\gamma_0+\gamma_1X_1+\gamma_2X_2+\gamma_3X_1^2+\gamma_4X_2^2),
\end{equation}
for some $\bmgamma=(\gamma_0,\gamma_1,\gamma_2,\gamma_3,\gamma_4)'$. The outcome variable follows the following relationship:
$$
\epn(Y|A,Z,\bmX)=h\left(-0.5+0.5X_1+0.2X_2-0.25X_1X_2+0.5X_2^2+A(0.5-0.1X_1)\right),
$$
where $h$ is the expit function for a binary outcome and the identity function for a continuous outcome. The true values of $(\mu_1,\mu_0,\delta)$ are $(0.5,0,0.5)$ in the continuous case and approximately $(0.600, 0.491, 0.109)$ in the binary case. The three parameters are estimated using the same six methods described earlier with modified working models that may be correct or incorrect. The correct PS model is given by \eqref{ps.mod.2}, and the incorrect one omits the quadratic terms. The correct OR model is a linear or logistic regression model with $(1, X_1, X_2, X_1X_2, X_2^2)$ as linear terms, and the incorrect one omits the interaction and quadratic terms.

Table \ref{sim.rst.1} shows the results of point estimation (empirical bias and standard deviation) for all six methods and three estimands. The results for estimating $\mu_1$ demonstrate that the augmented estimator $\widehat\mu_1^{\text{aug}}$ improves efficiency over $\overline Y_1$ without introducing bias, consistent with previous findings in the covariate adjustment literature. In what follows, we will focus on the estimation of $(\mu_0,\delta)$. The RCT-only and augmentation methods are both (virtually) unbiased in all cases, and the augmentation method is typically more efficient, at least when the OR model is correct. The unadjusted downweighting method is clearly biased with reduced variability as compared to the RCT-only method. The PS weighting method has negligible bias and reduced variability (as compared to the RCT-only method) when the PS model is correctly specified. When the PS model is mis-specified, the PS weighting method becomes visibly biased and its variability may or may not be lower than that of the RCT-only method. Likewise, the G-computation method is nearly unbiased when the OR model is correct and becomes visibly biased when the OR model is incorrect. The G-computation method does have lower variability than the RCT-only method and even the PS weighting method when the two model-based methods are compared on equal footing (i.e., when the OR and PS models are both correct or both incorrect). The weighted regression method is less susceptible to bias than the PS weighting and G-computation methods, showing a substantial bias only when the OR and PS models are both mis-specified. In terms of variability, the weighted regression method is generally similar to the G-computation method based on the same OR model, except that the weighted regression method appears to have lower variability when the OR model is incorrect and the PS model is correct.

For the three methods described in Section \ref{meth}, Table \ref{sim.rst.2} shows coverage proportions of 95\% Wald confidence intervals (CIs) based on analytical standard errors. For all configurations where the point estimator is consistent, the coverage proportions in Table 2 are close to the nominal level.

\section{Examples}\label{ex}


\subsection{Benign Prostate Hyperplasia (BPH)}\label{ex.bph}

Our first example concerns a trans-urethral microwave therapy (TUMT) for treating BPH. This example has been anonymized to protect confidentiality, and the anonymized dataset has been analyzed previously for illustrative purposes \citep{z09,z14}. The RCT in this example is a non-inferiority trial comparing an investigational TUMT device, say TUMT2, with an approved TUMT device, say TUMT1. The trial enrolled 200 male subjects over the age of 50 who had been diagnosed with BPH and had not been treated for it, with prostate size 20–50 cm$^3$ and American Urology Association Symptom Index (AUASI) at least 12. The AUASI score ranges between 0 and 35 with higher values indicating more severe symptoms. The subjects in the trial were randomized 1:1 to TUMT1 versus TUMT2. The primary efficacy endpoint was the mean decrease in AUASI from baseline to 6 months post-treatment. The observed mean decrease was 12.1 (0.7) in the TUMT2 arm and 13.6 (0.8) in the TUMT1 arm, with a difference of $-1.5$ (95\% CI: $-3.6$ to 0.6).

\begin{figure}
\centering
\includegraphics[width=0.6\textwidth]{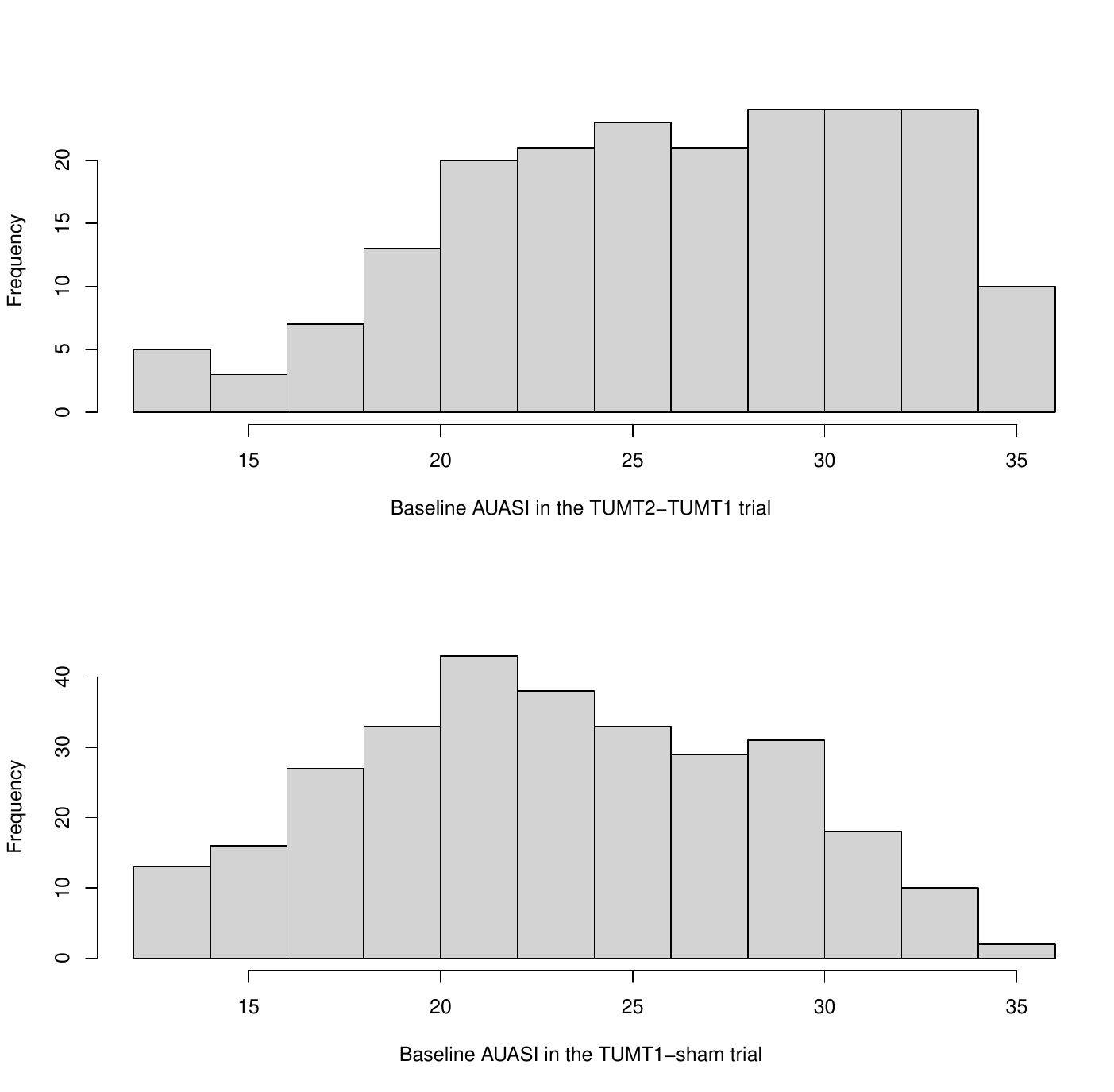}
\caption{Histograms of baseline AUASI scores in the two trials involved in the BPH example.}
\label{bph.hist}
\end{figure}

The external control arm in this example is the TUMT1 arm of a previous randomized trial comparing TUMT1 with a sham control, which \lq\lq treats" patients with the same device in the inactive mode. This trial enrolled 300 patients with similar inclusion-exclusion criteria and randomized them 2:1 to TUMT1 versus sham treatment. The observed mean decrease in AUASI over 6 months was 10.8 (0.5) in the TUMT1 arm. Note that the mean outcome in the external control arm is substantially and significantly ($p=0.002$) worse than that in the internal control arm, making it questionable to pool the two arms without any adjustment. This discrepancy may be related to the fact that the TUMT2-TUMT1 trial had higher baseline AUASI scores than the TUMT1-sham trial (see Figure \ref{bph.hist}) even though the two trials used the same entry criterion (baseline AUASI $\ge12$). It is plausible that patients with higher baseline AUASI scores tend to experience larger decreases regardless of the treatment received, a phenomenon known as \lq\lq regression to the mean". Thus, it makes sense to consider adjusting for baseline AUASI in the analysis of this hybrid control study.

The data are analyzed using the same six methods compared in Section \ref{sim} with the mean difference as the effect measure and with $w=0.25$ or 0.5 (corresponding to an effective sample size of 49.5 or 99 for the external control arm). The PS model is a logistic regression model given by \eqref{ps.mod.1} with $X$ being baseline AUASI. The OR model for both TUMT1 and TUMT2 is a linear regression model with $(1,X,X^2)$ as linear terms. For the PS-weighted estimates of $(\mu_0,\delta)$, standard errors are obtained using a nonparametric bootstrap procedure with 1,000 bootstrap samples. All other standard errors are derived from asymptotic variance formulas.

The results of this analysis are reported in Table \ref{bph.rst} in terms of point estimates and standard errors for estimating $(\mu_1,\mu_0,\delta)$. For all three parameters, augmentation produces similar point estimates and smaller standard errors in comparison to the RCT-only method. The unadjusted method produces notably different estimates of $(\mu_0,\delta)$ in favor of TUMT2, together with smaller standard errors, comparing again with the RCT-only method. Estimates of $(\mu_0,\delta)$ from the other three methods (PS weighting, G-computation and weighted regression) are generally similar to each other and fall between the RCT-only and unadjusted estimates. The smallest standard errors are produced by the G-computation and weighted regression methods. The impact of $w$ is quite large for the unadjusted method, relatively small for the PS weighting, G-computation and weighted regression methods, negligible for the augmentation method, and non-existent for the RCT-only method.

\subsection{Human Immunodeficiency Virus (HIV) Infection}\label{ex.hiv}

This second example concerns the efficacy of zidovudine (ZDV), a potent inhibitor of HIV replication, for treating HIV infection in asymptomatic patients with hereditary coagulation disorders. This question was evaluated in an RCT (ACTG036) that enrolled 193 patients from this population and randomized them 1:1 to ZDV or placebo \citep{m91}. The primary endpoint was the rate of treatment failure, defined as the occurrence of death, acquired immunodeficiency syndrome (AIDS), or advanced AIDS-related complex by 2 years of treatment. The observed failure rate was 4.5\% for ZDV and 7.4\% for placebo, with a difference of $-3.0$\% (95\% CI: $-9.8$\% to 3.9\%). This analysis suggests that ZDV may have a protective effect, but the supporting evidence is inconclusive.

In this example, the external control data come from the placebo arm of ACTG019, a randomized trial of ZDV versus placebo for treating HIV infection in asymptomatic patients with CD4 cell count lower than 500/cm$^2$ \citep{v90}. Patient-level data from both ACTG019 and ACTG036 are publicly available in the R package \texttt{hdbayes}. The observed failure rate was 8.9\% among the 404 patients in the placebo arm of ACTG019. It is not clear that the internal and external control data can be combined directly because the two trials had slightly different patient populations. Adjusting for baseline covariates would help to alleviate that concern. In the \texttt{hdbayes} version of trial data, there are three baseline covariates available: age, race (white or non-white), and CD4 cell count.

The data are analyzed using the same six methods employed for the BPH example with the failure rate difference as the effect measure and with $w=0.1$, 0.25 or 0.5 (corresponding to an effective sample size of 40.4, 101 or 202 for the external control arm). The PS model is a logistic regression model based on (age, race, $\sqrt{\text{CD4}}$) with interactions and quadratic terms. The OR model for placebo is a linear regression model based on (age, race, $\sqrt{\text{CD4}}$) with interactions. The OR model for ZDV is a linear regression model based on (age, race, $\sqrt{\text{CD4}}$). The OR models are limited in complexity by small numbers of failure events. Standard errors are obtained in the same manner as in the BPH example.

The results of this analysis are reported in Table \ref{hiv.rst}, where all three parameters are shown as percentages. The standard errors in Table \ref{hiv.rst} follow the same pattern as in Table \ref{bph.rst}. Compared to the RCT-only method, the unadjusted and PS weighting methods decrease the estimate of $\delta$ (in favor of ZDV) by an amount that increases with $w$, while the other three methods provide increased estimates of $\delta$. This example illustrates that the inclusion of external control data can reduce standard errors considerably (especially if the weighted regression or G-computation method is used), which in this case does not help to demonstrate the efficacy of ZDV.

\section{Discussion}\label{disc}

Clearly, the different methods for analyzing hybrid control studies provide different trade-offs between bias and variability. In light of the theoretical considerations in Section \ref{meth} and the simulation results in Section \ref{sim}, it makes sense to classify methods into three broad categories.

\begin{description}
\item{\bf Category 1} consists of methods that are guaranteed to be asymptotically unbiased under mild regularity conditions (excluding modeling assumptions). This category contains the RCT-only method, the augmentation method, and possibly some dynamic downweighting methods (whose asymptotic properties are not well understood by the authors). The augmentation method is generally more efficient than the RCT-only method. However, it is important to note that the efficiency gain comes from covariate adjustment and not from effective use of external control data. As discussed in Section \ref{aug}, there is no theoretical reason to expect $\widehat\mu_0^{\text{aug}}$ to become more efficient as a result of incorporating external control data or as the amount of external control data increases. The simulation results in Section \ref{sim} confirm that the augmentation method has limited capacity for efficiency improvement. It seems fair to conclude that the augmentation method uses external control data in a superficial way.
\item{\bf Category 2} consists of methods that are asymptotically unbiased under suitable modeling assumptions. This category contains the PS weighting method, the G-computation method, and the weighted regression method. All three methods can make effective use of external control data. Among the three methods, weighted regression is least susceptible to bias because of its double robustness property. We are not aware of general theoretical results on the variability of the three methods (as compared to each other). However, our simulation results suggest that G-computation tends to have less variability than PS weighting (when the OR and PS models are both correct or both incorrect) and that weighted regression is similar or slightly superior to G-computation in terms of variability. This comparison clearly favors weighted regression over the other two methods.
\item{\bf Category 3} consists of methods that are asymptotically unbiased only if the internal and external control arms happen to have the mean outcome. This category contains some static downweighting methods such as the unadjusted downweighting method considered here. Our simulation results demonstrate that the unadjusted method is able to introduce bias and reduce variability as compared to the RCT-only method. It is possible that the unadjusted method may sometimes provide a better bias-variance trade-off than a method from Category 2, especially if the latter is based on a mis-specified model. However, unlike the methods in Category 2, static downweighting methods do not provide an opportunity to reduce bias through careful modeling of observed data. We find this to be an unappealing feature of Category 3.
\end{description}

In practice, investigators have to consider the scientific and regulatory context of a hybrid control study---specifically the risk of bias and the potential benefit of incorporating external control data---in order to find the \lq\lq right" balance between bias and variability. If bias plays a dominant role in this trade-off, one may decide to choose a method from Category 1 such as the augmentation method. If one is willing to accept some bias in exchange for reduced variability, one might choose a method from Category 2 such as the weighted regression method. These recommendations are based on the currently available information about the few methods considered here; they are subject to revision as new information/methods become available. 

\section*{Acknowledgement}

Wei Liu's research was supported by the National Natural Science Foundation of China (Grant No.~12171121).

\pagebreak
\renewcommand{\baselinestretch}{0.91}
\begin{table}[htbp]
{\footnotesize
\caption{Simulation results for point estimation: empirical bias and standard deviation for estimating $(\mu_1,\mu_0,\delta)$ using six different estimation methods with correct or incorrect working models (see Section \ref{sim} for details).}\label{sim.rst.1}
\newcolumntype{d}{D{.}{.}{3}}
\begin{center}
\begin{tabular}{lllcdddcddd}
\hline
\hline
\multicolumn{1}{c}{Method}&\multicolumn{2}{c}{Working Model(s)}&&\multicolumn{3}{c}{Bias}&&\multicolumn{3}{c}{Std.~Deviation}\\
\cline{2-3}\cline{5-7}\cline{9-11}
&\multicolumn{1}{c}{OR}&\multicolumn{1}{c}{PS}&&\multicolumn{1}{c}{$\mu_1$}&\multicolumn{1}{c}{$\mu_0$}&\multicolumn{1}{c}{$\delta$}&&\multicolumn{1}{c}{$\mu_1$}&\multicolumn{1}{c}{$\mu_0$}&\multicolumn{1}{c}{$\delta$}\\
\hline
\multicolumn{11}{c}{one covariate, binary outcome}\\
RCT-only&&&&0.000&0.001&-0.001&&0.049&0.071&0.085\\
Augmentation&correct&&&0.000&0.001&-0.001&&0.048&0.069&0.082\\
Augmentation&incorrect&&&-0.001&0.000&-0.001&&0.049&0.071&0.085\\
Unadjusted&&&&0.000&0.044&-0.044&&0.049&0.043&0.064\\
PS weighting&&correct&&0.000&0.001&-0.001&&0.049&0.046&0.066\\
PS weighting&&incorrect&&0.000&0.050&-0.050&&0.049&0.044&0.065\\
G-computation&correct&&&0.000&0.001&-0.001&&0.048&0.046&0.065\\
G-computation&incorrect&&&-0.001&0.041&-0.042&&0.049&0.045&0.066\\
Wtd.~regression&correct&correct&&0.000&0.000&0.000&&0.048&0.045&0.065\\
Wtd.~regression&correct&incorrect&&0.000&0.000&0.000&&0.048&0.046&0.065\\
Wtd.~regression&incorrect&correct&&0.000&-0.001&0.001&&0.049&0.046&0.067\\
Wtd.~regression&incorrect&incorrect&&0.000&0.041&-0.042&&0.049&0.045&0.066\\
\hline
\multicolumn{11}{c}{one covariate, continuous outcome}\\
RCT-only&&&&0.002&-0.001&0.002&&0.124&0.180&0.219\\
Augmentation&correct&&&0.001&0.001&0.000&&0.117&0.155&0.173\\
Augmentation&incorrect&&&-0.002&-0.004&0.002&&0.124&0.183&0.219\\
Unadjusted&&&&0.002&0.300&-0.299&&0.124&0.131&0.180\\
PS weighting&&correct&&0.002&-0.001&0.003&&0.124&0.119&0.161\\
PS weighting&&incorrect&&0.002&0.359&-0.357&&0.124&0.183&0.221\\
G-computation&correct&&&0.001&0.000&0.000&&0.117&0.110&0.135\\
G-computation&incorrect&&&-0.002&0.269&-0.271&&0.124&0.140&0.186\\
Wtd.~regression&correct&correct&&0.001&0.000&0.000&&0.117&0.110&0.134\\
Wtd.~regression&correct&incorrect&&0.001&0.000&0.001&&0.117&0.111&0.135\\
Wtd.~regression&incorrect&correct&&-0.002&-0.005&0.003&&0.124&0.117&0.158\\
Wtd.~regression&incorrect&incorrect&&-0.002&0.292&-0.294&&0.124&0.140&0.184\\
\hline
\multicolumn{11}{c}{two covariates, binary outcome}\\
RCT-only&&&&0.001&0.000&0.000&&0.049&0.072&0.088\\
Augmentation&correct&&&0.001&0.000&0.000&&0.048&0.068&0.082\\
Augmentation&incorrect&&&0.000&0.000&0.000&&0.049&0.071&0.086\\
Unadjusted&&&&0.001&0.015&-0.014&&0.049&0.044&0.066\\
PS weighting&&correct&&0.001&0.000&0.001&&0.049&0.050&0.069\\
PS weighting&&incorrect&&0.001&0.057&-0.057&&0.049&0.048&0.068\\
G-computation&correct&&&0.001&0.000&0.000&&0.048&0.048&0.066\\
G-computation&incorrect&&&0.000&0.039&-0.039&&0.049&0.047&0.067\\
Wtd.~regression&correct&correct&&-0.001&0.001&-0.001&&0.048&0.049&0.065\\
Wtd.~regression&correct&incorrect&&-0.001&0.001&-0.002&&0.048&0.049&0.065\\
Wtd.~regression&incorrect&correct&&-0.001&-0.001&-0.001&&0.048&0.049&0.067\\
Wtd.~regression&incorrect&incorrect&&-0.001&0.040&-0.042&&0.048&0.046&0.065\\
\hline
\multicolumn{11}{c}{two covariates, continuous outcome}\\
RCT-only&&&&0.000&0.001&-0.001&&0.133&0.193&0.234\\
Augmentation&correct&&&0.000&0.001&-0.001&&0.123&0.159&0.172\\
Augmentation&incorrect&&&-0.004&-0.001&-0.002&&0.131&0.185&0.219\\
Unadjusted&&&&0.000&0.171&-0.171&&0.133&0.147&0.199\\
PS weighting&&correct&&0.000&-0.002&0.002&&0.133&0.134&0.175\\
PS weighting&&incorrect&&0.000&0.401&-0.401&&0.133&0.198&0.234\\
G-computation&correct&&&0.000&0.000&0.000&&0.123&0.119&0.136\\
G-computation&incorrect&&&-0.004&0.271&-0.275&&0.131&0.150&0.191\\
Wtd.~regression&correct&correct&&0.000&0.000&0.000&&0.123&0.122&0.138\\
Wtd.~regression&correct&incorrect&&0.000&0.000&0.000&&0.123&0.120&0.137\\
Wtd.~regression&incorrect&correct&&-0.004&-0.005&0.001&&0.131&0.129&0.165\\
Wtd.~regression&incorrect&incorrect&&-0.004&0.295&-0.298&&0.131&0.156&0.195\\
\hline
\end{tabular}
\end{center}
}
\end{table}

\pagebreak
\renewcommand{\baselinestretch}{1.0}
\begin{table}[htbp]
\caption{Simulation results for interval estimation: empirical coverage proportions of 95\% Wald confidence intervals for $(\mu_1,\mu_0,\delta)$, obtained using the three estimation methods described in Section \ref{meth} with correct or incorrect working models (see Section \ref{sim} for details).}\label{sim.rst.2}
\newcolumntype{d}{D{.}{.}{3}}
\begin{center}
\begin{tabular}{lllcddd}
\hline
\hline
\multicolumn{1}{c}{Method}&\multicolumn{2}{c}{Working Model(s)}&&\multicolumn{3}{c}{Coverage Proportion}\\
\cline{2-3}\cline{5-7}
&\multicolumn{1}{c}{OR}&\multicolumn{1}{c}{PS}&&\multicolumn{1}{c}{$\mu_1$}&\multicolumn{1}{c}{$\mu_0$}&\multicolumn{1}{c}{$\delta$}\\
\hline
\multicolumn{7}{c}{one covariate, binary outcome}\\
Augmentation&correct&&&0.944&0.940&0.947\\
Augmentation&incorrect&&&0.947&0.940&0.947\\
G-computation&correct&&&0.944&0.947&0.950\\
G-computation&incorrect&&&0.947&0.845&0.904\\
Wtd.~regression&correct&correct&&0.944&0.948&0.947\\
Wtd.~regression&correct&incorrect&&0.944&0.944&0.947\\
Wtd.~regression&incorrect&correct&&0.945&0.944&0.948\\
Wtd.~regression&incorrect&incorrect&&0.945&0.846&0.903\\
\hline
\multicolumn{7}{c}{one covariate, continuous outcome}\\
Augmentation&correct&&&0.946&0.946&0.945\\
Augmentation&incorrect&&&0.946&0.940&0.946\\
G-computation&correct&&&0.946&0.948&0.948\\
G-computation&incorrect&&&0.946&0.507&0.687\\
Wtd.~regression&correct&correct&&0.946&0.949&0.947\\
Wtd.~regression&correct&incorrect&&0.946&0.948&0.949\\
Wtd.~regression&incorrect&correct&&0.946&0.931&0.957\\
Wtd.~regression&incorrect&incorrect&&0.946&0.428&0.639\\
\hline
\multicolumn{7}{c}{two covariates, binary outcome}\\
Augmentation&correct&&&0.945&0.940&0.939\\
Augmentation&incorrect&&&0.946&0.940&0.942\\
G-computation&correct&&&0.945&0.945&0.943\\
G-computation&incorrect&&&0.946&0.858&0.905\\
Wtd.~regression&correct&correct&&0.944&0.946&0.950\\
Wtd.~regression&correct&incorrect&&0.944&0.945&0.951\\
Wtd.~regression&incorrect&correct&&0.945&0.943&0.950\\
Wtd.~regression&incorrect&incorrect&&0.945&0.851&0.906\\
\hline
\multicolumn{7}{c}{two covariates, continuous outcome}\\
Augmentation&correct&&&0.945&0.944&0.947\\
Augmentation&incorrect&&&0.946&0.945&0.947\\
G-computation&correct&&&0.945&0.952&0.949\\
G-computation&incorrect&&&0.946&0.552&0.696\\
Wtd.~regression&correct&correct&&0.945&0.951&0.951\\
Wtd.~regression&correct&incorrect&&0.945&0.952&0.949\\
Wtd.~regression&incorrect&correct&&0.946&0.938&0.957\\
Wtd.~regression&incorrect&incorrect&&0.946&0.500&0.651\\
\hline
\end{tabular}
\end{center}
\end{table}

\pagebreak
\renewcommand{\baselinestretch}{1.0}
\begin{table}[htbp]
\caption{Analysis of BPH example data: point estimates (standard errors) of $(\mu_1,\mu_0,\delta)$ from six different estimation methods with $w=0.25$ or 0.5 (see Section \ref{ex.bph} for details).}\label{bph.rst}
\newcolumntype{d}{D{.}{.}{2}}
\begin{center}
\begin{tabular}{llrrr}
\hline
\hline
\multicolumn{1}{c}{Method}&\multicolumn{1}{c}{$w$}&\multicolumn{3}{c}{Pt.~Est.~(Std.~Err.)}\\
\cline{3-5}
\multicolumn{2}{c}{}&\multicolumn{1}{c}{$\mu_1$}&\multicolumn{1}{c}{$\mu_2$}&\multicolumn{1}{c}{$\delta$}\\
\hline
RCT-only&&12.1 (0.7)&13.6 (0.8)&$-1.5$ (1.1)\\
\hline
Augmentation&0.25&12.1 (0.6)&13.4 (0.7)&$-1.4$ (0.9)\\
Unadjusted&0.25&12.1 (0.7)&12.7 (0.6)&$-0.5$ (0.9)\\
PS weighting&0.25&12.1 (0.7)&13.1 (0.6)&$-1.1$ (0.9)\\
G-computation&0.25&12.1 (0.6)&13.3 (0.6)&$-1.3$ (0.7)\\
Wtd.~regression&0.25&12.1 (0.6)&13.2 (0.6)&$-1.1$ (0.7)\\
\hline
Augmentation&0.5&12.1 (0.6)&13.4 (0.7)&$-1.4$ (0.9)\\
Unadjusted&0.5&12.1 (0.7)&12.2 (0.5)&0.0 (0.9)\\
PS weighting&0.5&12.1 (0.7)&13.1 (0.6)&$-1.0$ (0.8)\\
G-computation&0.5&12.1 (0.6)&13.2 (0.5)&$-1.2$ (0.7)\\
Wtd.~regression&0.5&12.1 (0.6)&13.0 (0.5)&$-1.0$ (0.7)\\
\hline
\end{tabular}
\end{center}
\end{table}

\renewcommand{\baselinestretch}{1.0}
\begin{table}[htbp]
\caption{Analysis of HIV example data: point estimates (standard errors) of $(\mu_1,\mu_0,\delta)$ from six different estimation methods with $w=0.1$, 0.25 or 0.5 (see Section \ref{ex.hiv} for details).}\label{hiv.rst}
\newcolumntype{d}{D{.}{.}{2}}
\begin{center}
\begin{tabular}{llrrr}
\hline
\hline
\multicolumn{1}{c}{Method}&\multicolumn{1}{c}{$w$}&\multicolumn{3}{c}{Pt.~Est.~(Std.~Err.)}\\
\cline{3-5}
\multicolumn{2}{c}{}&\multicolumn{1}{c}{$\mu_1$ (\%)}&\multicolumn{1}{c}{$\mu_2$ (\%)}&\multicolumn{1}{c}{$\delta$ (\%)}\\
\hline
RCT-only&&4.5 (2.2)&7.4 (2.7)&$-3.0$ (3.5)\\
\hline
Augmentation&0.1&6.3 (2.0)&6.6 (2.5)&$-0.3$ (2.9)\\
Unadjusted&0.1&4.5 (2.2)&7.9 (2.0)&$-3.4$ (2.9)\\
PS weighting&0.1&4.5 (2.2)&7.6 (2.1)&$-3.1$ (3.1)\\
G-computation&0.1&6.3 (2.0)&7.6 (1.9)&$-1.3$ (2.5)\\
Wtd.~regression&0.1&6.3 (2.0)&6.8 (1.8)&$-0.5$ (2.3)\\
\hline
Augmentation&0.25&6.3 (2.0)&6.6 (2.5)&$-0.3$ (2.9)\\
Unadjusted&0.25&4.5 (2.2)&8.2 (1.5)&$-3.7$ (2.7)\\
PS weighting&0.25&4.5 (2.2)&7.8 (1.8)&$-3.3$ (2.7)\\
G-computation&0.25&6.3 (2.0)&8.1 (1.8)&$-1.9$ (2.3)\\
Wtd.~regression&0.25&6.3 (2.0)&7.1 (1.5)&$-0.8$ (2.2)\\
\hline
Augmentation&0.5&6.3 (2.0)&6.6 (2.5)&$-0.3$ (3.0)\\
Unadjusted&0.5&4.5 (2.2)&8.4 (1.3)&$-4.0$ (2.6)\\
PS weighting&0.5&4.5 (2.2)&7.9 (1.8)&$-3.4$ (2.7)\\
G-computation&0.5&6.3 (2.0)&8.4 (1.6)&$-2.1$ (2.3)\\
Wtd.~regression&0.5&6.3 (2.0)&7.4 (1.4)&$-1.1$ (2.1)\\
\hline
\end{tabular}
\end{center}
\end{table}

\end{document}